\documentclass[
  twoside,
  floatfix,
  twocolumn,
  aps,
  10pt,
  prb,
  floatfix,
  showpacs,
  citeautoscript,
  superscriptaddress,
  longbibliography,
]{revtex4-2}

\usepackage{amsmath}
\usepackage{amssymb}
\usepackage{dcolumn}
\usepackage[acronym]{glossaries}
\usepackage{graphicx}
\usepackage{siunitx}
\usepackage{tabularx}
\usepackage{multirow}
\usepackage[dvipsnames]{xcolor}
\usepackage[colorlinks]{hyperref}
\hypersetup{
    linkcolor=Black,
    citecolor=Black,
    filecolor=Black,
    urlcolor=Black,
}

\graphicspath{{submission-v1/images/}}

\setacronymstyle{long-short}
\newacronym{mc}{MC}{Monte Carlo}
\newacronym{sh}{SH}{spherical harmonic}
\newacronym{saxs}{SAXS}{small-angle X-ray scattering}
\newacronym{saxstt}{SAXSTT}{small-angle X-ray scattering tensor tomography}
\newacronym{sigtt}{SIGTT}{spherical integral geometric tensor tomography}
\newacronym{rsm}{RSM}{reciprocal space map}

\DeclareMathOperator*{\argmin}{argmin}

\newcommand{\chalmersphys}{
  Chalmers University of Technology,
  Department of Physics,
  Gothenburg, Sweden
}
\newcommand{\affilpsi}{
    Paul Scherrer Institute (PSI), 
    Villigen,
    Switzerland
}
\newcommand{\affilepfl}{
    École Polytechnique Fédérale de Lausanne (EPFL),
    Lausanne,
    Switzerland
}  

\makeatletter
\newcommand\setcurrentname[1]{\def\@currentlabelname{#1}}
\makeatother
\begin{document}
\title{
    Small-angle scattering tensor tomography algorithm \\
    for robust reconstruction of complex textures
}

\author{Leonard C. Nielsen}
\email{leoniel@chalmers.se}
\affiliation{\chalmersphys}
\author{Paul Erhart}
\affiliation{\chalmersphys}
\author{Manuel Guizar-Sicairos}
\affiliation{\affilpsi}
\affiliation{\affilepfl}
\author{Marianne Liebi}
\email[Corresponding author. ]{marianne.liebi@psi.ch}
\affiliation{\chalmersphys}
\affiliation{\affilpsi}
\affiliation{\affilepfl}

\begin{abstract}
The development of small-angle scattering tensor tomography has enabled the study of anisotropic nanostructures in a volume-resolved manner.
It is of great value to have reconstruction methods that can handle many different nanostructural symmetries.
For such a method to be employed by researchers from a wide range of backgrounds, it is crucial that its reliance on prior knowledge about the system is minimized, and that it is robust under various conditions.
Here, we present a method employing band-limited spherical functions to enable the reconstruction of reciprocal space maps of a wide variety of nanostructures.
This method has been thoroughly tested and compared to existing methods in its ability to retrieve known reciprocal space maps, as well as its robustness to changes in initial conditions, using both simulations and experimental data.
The anchoring of this method in a framework of integral geometry and linear algebra highlights its possibilities and limitations.
\end{abstract}

\maketitle


\Gls{saxs} probes the nanometer-scale variations in the electron density of materials averaged over areas of scales typically in the range of \SI{1x1}{\micro\meter} to \SI{1000x1000}{\micro\meter}, depending on the size of the X-ray beam.
The data from a \gls{saxs} experiment carries information about the nanostructure of a sample, including characteristic length scales and orientation, and has been used to study numerous materials \cite{fratzl_ctr_1996, marios_rst_2016, licht_jac_1999}.
Scanning \gls{saxs} can be performed across a sample to yield a two-dimensional map, with each scanned pixel associated with a corresponding two-dimensional cut through the reciprocal space map \cite{fratzl_jac_1997, pabisch_book_2013, paris_bio_2008}.
Since a \gls{saxs} measurement with an area detector gives two-dimensional data, measurements must be performed at several angles to obtain three-dimensional data from a sample \cite{liu_jp_2010, seidel_bbn_2012, marios_rst_2016}.
Rotating a three-dimensional sample around a single axis, standard tomographic reconstruction can be used in cases where the sample scattering is isotropic or when the scattering is symmetric about the axis of rotation \cite{feldkamp_pss_2009, schroer_apl_2006, stribeck_jps_2008, alvarez_jac_2012, jensen_ni_2011}.

With the use of a tilt angle in addition to rotation, recent works by Schaff \textit{et al.}\ (2015) \cite{schaff_nature_2015}, Liebi \textit{et al.}\ (2015, 2018) \cite{liebi_nat_2015, liebi_aca_2018} and Gao \textit{et al.}\ (2018) \cite{gao_aca_2019} present tomographic methods for the reconstruction of the three-dimensional reciprocal space map using scanning \gls{saxs} projections, or \gls{saxstt}.
Whereas Schaff \textit{et al.}\ (2015) \cite{schaff_nature_2015} demonstrated a two-step procedure of first reconstructing the reciprocal space map without any data-reducing assumptions and then analyzing this reconstruction to find orientations, the approaches of Liebi \textit{et al.}\ (2015, 2018) \cite{liebi_nat_2015, liebi_aca_2018} (referred to as the Spherical Harmonic or SH method) and Gao \textit{et al}.\ (2018)\cite{gao_aca_2019} (referred to as the Iterative Reconstruction or IR method) use reduced models of the reciprocal space maps.
The SH method \cite{liebi_nat_2015, liebi_aca_2018} models the reciprocal space map of each voxel using squared band-limited spherical functions expressed in spherical harmonics.
The demonstrations of the method in \cite{liebi_nat_2015, liebi_aca_2018} used a reduced model, employing only zonal harmonics (spherical harmonics symmetric about an axis of rotation) and two Euler angles, with the angles parameterising the main nanostructural orientation.
The IR method \cite{gao_aca_2019} uses the symmetric rank-2 tensor as the basis for its model.
The work of Gao \textit{et al}.\ (2018) \cite{gao_aca_2019} primarily describes the reconstruction of a proposed orientation distribution function derived heuristically for fiber scattering.
The model of the scattering as derived from orientation distribution functions is based on different assumptions from those of the model of the scattering as the sum of contributions from cuts through the reciprocal space map.
The two models are not in general compatible, as the proposed orientation distribution function does not possess the rotational invariance necessary to map it to spherical reciprocal space map functions.
However, as noted in a footnote in Gao \textit{et al}.\ (2018) \cite{gao_aca_2019}, it is also possible to configure the IR algorithm to reconstruct the reciprocal space map directly, which is the way IR was employed in this work.

The use of a complete band-limited basis of even-ordered spherical harmonics described in Liebi \textit{et al.}\ (2018) has to the best of our knowledge not been implemented or tested for \gls{saxstt} as of the writing of this work.
Moreover, the approach in \cite{liebi_aca_2018, liebi_nat_2015} requires fitting the measured reciprocal space map to sums of squared polynomials, which is a difficult class of optimization problems to solve \cite{8263706}, as it results in a non-linear system of equations.
Here, we present \gls{sigtt} as an improvement on the complete basis approach proposed by Liebi \textit{et al}.\ (2015, 2018).
\gls{sigtt} eliminates the squaring of the polynomial which was used in the SH approach, resulting in a linear system, and its implementation does not rely on Euler angles.
\section*{Results}
\label{sect:results}

\subsection*{Theory}
\label{sect:theory}

The equation for the measured reciprocal space map of \gls{saxs} may be written as

\begin{align}
    \textsc{RSM}(\mathbf{q})
    &=
    \iiint dV \big[\Tilde{\rho}(\mathbf{r}) e^{-i \mathbf{q} \cdot \mathbf{r}}],
    \label{eq:reciprocal_space_map}
\end{align}

where $\tilde{\rho}(\mathbf{r})$ is the auto-correlation function of the electron density of the probed volume, $\mathbf{r}$ is the position, and $\mathbf{q}$ is the reciprocal space vector.
Consider this integral over a region of space (a voxel) and at a fixed $\Vert\mathbf{q}\Vert$.
Then, $\textsc{RSM}(\mathbf{q})$ reduces to $\textsc{RSM}(\theta,\phi)$, a function on the unit sphere, which may by theorem be represented by an infinite series of spherical harmonics \cite{volker_2019_spherical}.
As discussed in previous work by Liebi \textit{et al.} (2018) \cite{liebi_aca_2018}, the summation is reduced to the even orders as a result of Friedel symmetry.
The measured reciprocal space map at a single $\mathbf{q}$ and at a particular position $\mathbf{r}$ in space may then be written $\textsc{RSM}(\mathbf{r}, \theta, \phi)$ and expanded in spherical harmonics as

\begin{align}
    \label{eq:voxel_model}
    \textsc{RSM}(\mathbf{r},\theta,\phi) = \sum_{\ell=0,2,\hdots}^{\infty}\sum_{m=-\ell}^{\ell} a_m^\ell(\mathbf{r}) \hat{Y}^\ell_m(\theta,\phi),
\end{align}

where $\theta$ and $\phi$ are the polar and azimuthal angles of the reciprocal space map, $\hat{Y}^\ell_m(\theta,\phi)$ is the real-valued spherical harmonic basis function of order $\ell$ and degree $m$, and $a_m^\ell(\mathbf{r})$ is the coefficient of that basis function at position $\mathbf{r}$.
Note that the summation over $\ell$ in equation~\eqref{eq:johntransform} only includes even terms, as indicated by the subscript of the summation sign.




The following projection model is based on the discrete model in Liebi \textit{et al.}\ (2018), cast in terms of line integrals within the sample coordinate system, which map it onto a projection space.
The projection space is spanned by four coordinates $(j,k,\alpha,\beta)$, which can be mapped to experimental parameters.
The linear coordinates, $(j,k)$, map to the vertical and horizontal positioning of the sample during scanning \gls{saxs}.
The angular coordinates $(\alpha,\beta)$ map to rotations of the sample about laboratory axes orthogonal to the beam direction during the measurement.
Rotations about axes which are not orthogonal to the beam direction can be handled by decomposition into beam-orthogonal rotations that map to $(\alpha, \beta)$, and beam-parallel rotations which map to transforms of $(j, k)$.
The projection of a scalar function $f(\mathbf{r})$ from three to two dimensions at an arbitrary angle for a narrow beam is described by the John transform, also known as the X-ray transform, which may be expressed

\begin{align*}
\mathbb{P}[f](j,k,\alpha,\beta)
=
\int_{-\infty}^{\infty} f(\mathbf{v}(j,k,\alpha,\beta) + s\mathbf{u}(\alpha,\beta)) \mathop{ds},
\end{align*}

where $\alpha$ and $\beta$ are the azimuthal and polar angles, respectively, with respect to a fixed plane in the sample coordinate system, for a line of integration which intersects with the system's origin.
Then, $j$ and $k$ give the line's offset from the origin in the plane of projection, which is orthogonal to the line's direction.
In this parameterization, $\mathbf{v}$ represents the position of the beam in the plane of projection, whereas $\mathbf{u}$ represents its direction.
See Supplementary Note 1 for more details on how to map the sample coordinate system to an experimental coordinate system.
To keep our equations compact, we will use the shorthand notation

\begin{align}
\label{eq:upsilon}
\overline{\upsilon} \equiv (\upsilon_1,\upsilon_2,\upsilon_3,\upsilon_4) \equiv (j,k,\alpha,\beta),
\end{align}

such that $\overline{\upsilon}$ represents simultaneously a line of integration in three-dimensional space and a measured point in projection space.
By inserting equation~\eqref{eq:voxel_model} into the John transform $\mathbb{P}[f]$, we obtain an expression for the projection of the spherical harmonics from three to two dimensions using two projection angles:

\begin{align}
\mathbb{P}[\textsc{RSM}]&(\theta,\phi, \overline{\upsilon})
= \sum_{\ell=0,2,\ldots}^{\infty}\sum_{m=-\ell}^{\ell} \hat{Y}^\ell_m(\theta,\phi) A_{m}^\ell(\overline{\upsilon})
\label{eq:johntransform}
\end{align}

with

\begin{align*}
A_{m}^\ell(\overline{\upsilon})
=
\int_{-\infty}^{\infty} a_m^\ell
(
    \mathbf{v}(\overline{\upsilon}) + s\mathbf{u}(\upsilon_3,\upsilon_4)
) \mathop{ds}
\end{align*}

From equation~\eqref{eq:johntransform} it is apparent that both the reciprocal space map at a particular point in space and the projection of it may be represented using an even-ordered spherical harmonic expansion with a one-to-one correspondence between the representations.
In other words, each order and degree in the projected harmonic can be regarded as the projection of a single order and degree in the harmonics distributed in space.
However, small-angle scattering does not permit us to probe the entirety of the reciprocal space map at a single projection angle; instead, it probes a set of points which lie approximately on a great circle spanned by the set of unit vectors orthogonal to the direction of the beam.
We describe this using a parametric curve $C(\varphi, \alpha, \beta)$, where for fixed $\alpha$ and $\beta$, $C(\varphi)$ is a great circle orthogonal to the direction of $\mathbf{u}$.


We can then write the model (which we will denote $\mathcal{I}$) for a single measuremed reciprocal space map as

\begin{align}
\mathcal{I}(\varphi,\overline{\upsilon}) 
&= \sum_{\ell=0,2}^{\infty}\sum_{m=-\ell}^{\ell} \hat{Y}^\ell_m(C(\varphi,\upsilon_3,\upsilon_4)) A_{m}^\ell(\overline{\upsilon}),
\label{eq:single_measurement}
\end{align}

which completes the forward model of \gls{sigtt}.
Anisotropic scattering signals have previously been modeled using line integrals of spherical polynomials by Wieczorek \textit{et al.}\ (2016) for dark-field tomography \cite{darkfield_prl_2016}.
The key difference of \gls{sigtt} from the spherical harmonic model described in Liebi \textit{et al.}\ (2018) \cite{liebi_aca_2018} is the linearity of the system, which is achieved by not squaring the spherical harmonics.
Unlike the implementation demonstrated in Liebi \textit{et al.}\ (2015, 2018) \cite{liebi_aca_2018, liebi_nat_2015}, \gls{sigtt} does not employ a local coordinate system for the reciprocal space map in each voxel.

\begin{figure*}
    \includegraphics[width=\linewidth]{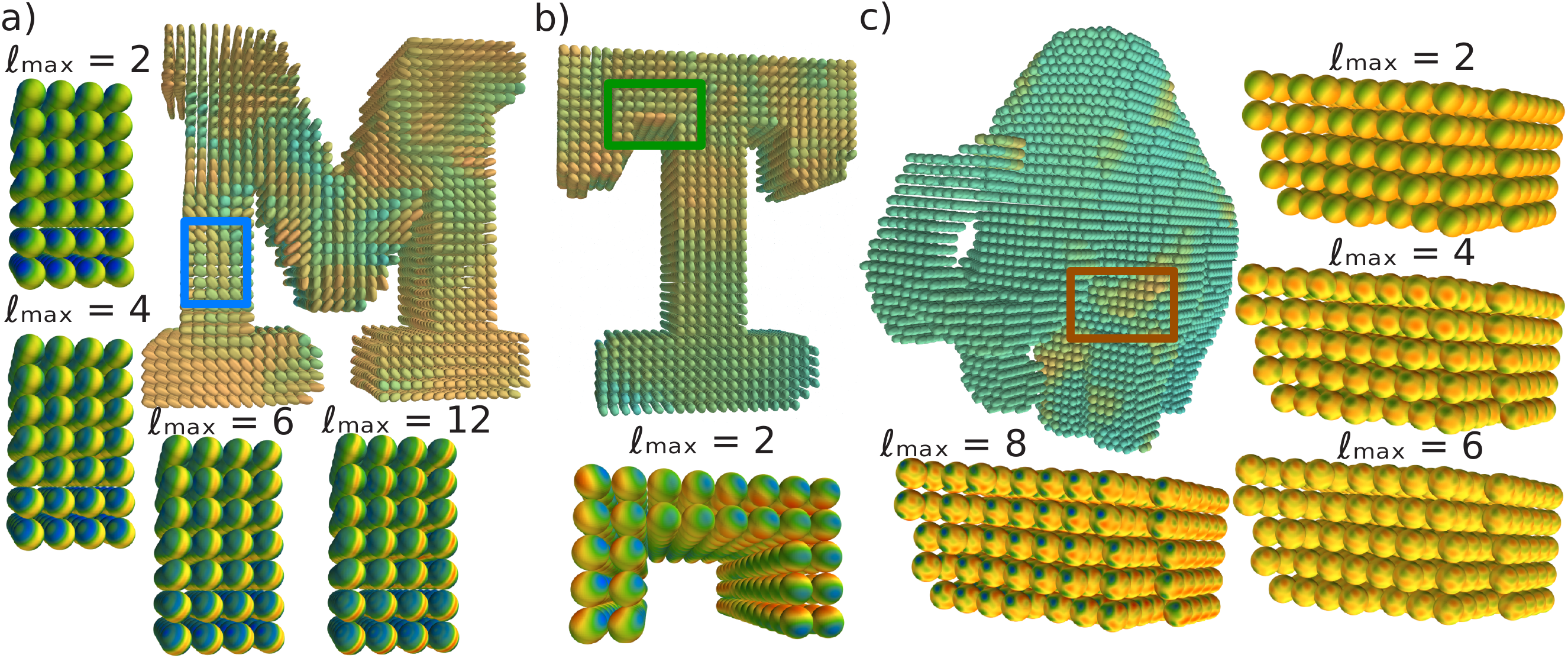}
  \caption{
    \textbf{Superquadric glyph render of simulations.}
    \textbf{a)} Sample ``M'' with its zonally symmetric reciprocal space maps from one region (blue square), truncated at $\ell = 2,\ 4,\ 6,\ 12$ respectively.
    \textbf{b)} Sample ``T'' with its rank-2 tensor reciprocal space maps from one region (green square).
    \textbf{c)} Sample ``mammoth'' with its unrestricted $\ell_\text{max} = 8$ reciprocal space maps from one region (orange square), truncated at $\ell = 2,\ 4,\ 6,\ 8$ respectively.
  }
  \label{fig:samples}
\end{figure*}
These changes result in the preservation of the orthogonality of each component of the per-voxel model, and simplifies gradient calculations.
See Supplementary Note 2 for how the \gls{sigtt} representation maps to Cartesian tensors, as in the model used by IR.
The inverse problem of obtaining $a^\ell_m(\mathbf{r})$ is solved by using a regularized least-squares approach (\nameref{sect:methods}).

\subsection*{Simulations}
\label{sect:simulations}
In order to compare and evaluate the different methods, a simulation framework was developed (\nameref{sect:methods}).
In total, three simulations were created, labelled ``M'', ``T'', and ``mammoth'', see \autoref{fig:samples} for an overview.
Sample ``M'' is intended to provide a simulated reciprocal space map with an intensity distribution similar to what is assumed by SH.
This is done by constraining its simulated reciprocal space map to be approximately described by zonal harmonics up to $\ell_\text{max} = 12$.
In the simulation of ``T'', the reciprocal space map is described entirely by symmetric rank-2 tensors, which is the model employed by IR.
Finally, the simulation of ``mammoth'' employed spherical harmonics up to $\ell_\text{max} = 8$, with a weak $\ell = 2$ component, to model reciprocal space maps with more complicated symmetries.
The reconstructions of each method are compared to the simulated samples by calculating the squared Pearson correlation coefficient $R^2$ (equation~\eqref{eq:corr}) of the simulated and reconstructed reciprocal space maps on a voxel-by-voxel basis.

\begin{figure*}
\includegraphics[width=\linewidth]{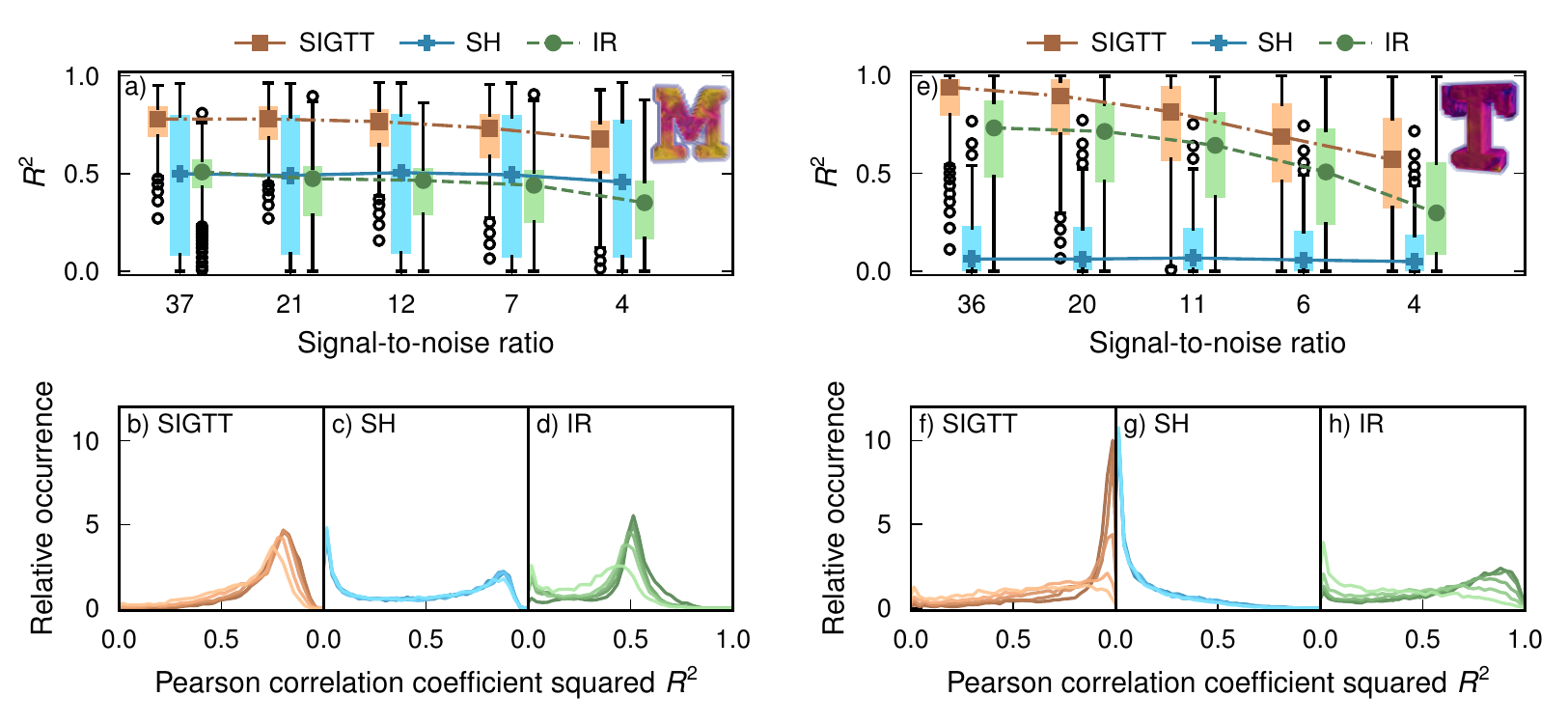}
\caption{
    \textbf{Correlations for samples ``M'' and ``T''.}
    \textbf{a)--d)} Comparison between SIGTT, SH and IR for the sample ``M''. \textbf{e)--h)} Comparison for sample ``T''. The image insets show volume renders of each sample. Panels a) and e) show box plots of $R^2$, the correlations with the reciprocal space maps of the simulated sample, as defined in equation~\eqref{eq:corr}, with lines and symbols indicating the median in each box plot. Outlier dots each represent $100$ voxels. 
    Panels b)--d) and f)--h) show how the correlation coefficient is distributed for each of the three methods. The signal-to-noise ratio goes from 37 (darkest lines) down to 4 (lightest lines) in b)--d), and from 36 (darkest lines) down to 4 (lightest lines) in f)--h).}
\label{fig:zonal_and_r2}
\end{figure*}
In \autoref{fig:zonal_and_r2}, the results of reconstructing simulated data for samples ``M'' and ``T'' are shown.
See \nameref{sect:methods} for details on the box plots.
The subplots labelled a)--d) show the performance of \gls{sigtt}, SH and IR in reconstructing sample ``M'', which consists of approximately zonal harmonics, possessing a great circle of intensity.
\gls{sigtt} performs the best in the comparison, with a peak correlation centered around $R^2 = 0.8$ at the highest SNR, decaying down to about $0.75$ at the lowest SNR, with a greater interquartile range.
The best possible performance of \gls{sigtt} in this comparison is constrained both by the fact that the chosen discretization of the reciprocal space map permits the fitting of harmonics only up to $\ell = 6$, whereas the sample contains harmonics up to $\ell = 12$.
The reconstruction is also affected by the missing wedge problem, a common limitation in tomography when only a subset of projection space is be sampled.
The performance of SH is almost unaffected by the SNR, with its correlation centered around $R^2 = 0.5$ with a large interquartile range.
In the lower middle plot, a peak can be seen both around $0.8$ and around $0$, suggesting a mixture of good and poor performance across the sample volume.
The trend in the performance of IR is more similar to that of \gls{sigtt} in that its correlation decays and its interquartile range increases with the decrease in SNR.
However, its maximum possible correlation to most of the sample is bounded at around $0.5$ by the fact that it can only correlate to the $\ell = 2$ component of the model, due to being restricted to the symmetric rank-2 tensor.
\begin{figure}
    \includegraphics[width=\columnwidth]{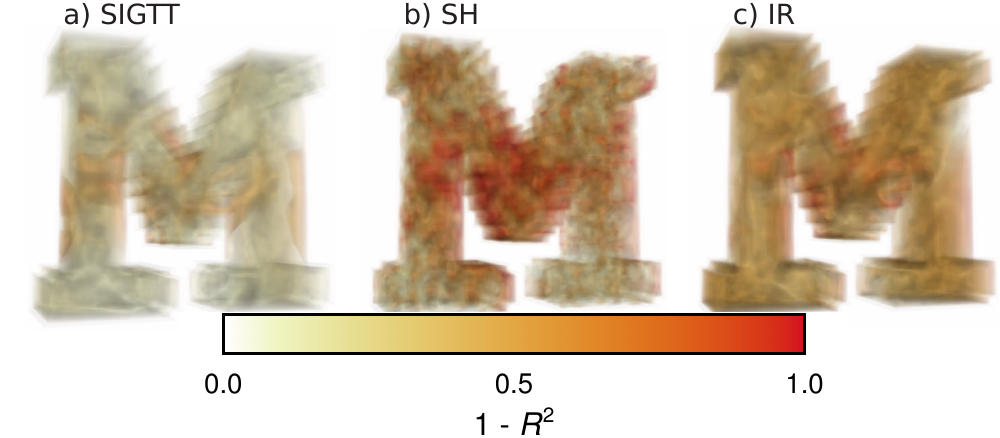}
  \caption{
    \textbf{Volume renders of errors for reconstructions of ``M''.} \textbf{a)} Errors for SIGTT.
    \textbf{b)} Errors for SH.
    \textbf{c)} Errors for IR.
    The error is defined as $1 - R^2$, where $R^2$ is given by equation~\eqref{eq:corr}. Larger errors are rendered with greater opacity and are thus visible even if they are in the interior.
  }
  \label{fig:m_errors}
\end{figure}
\begin{figure*}
    \includegraphics[width=\linewidth]{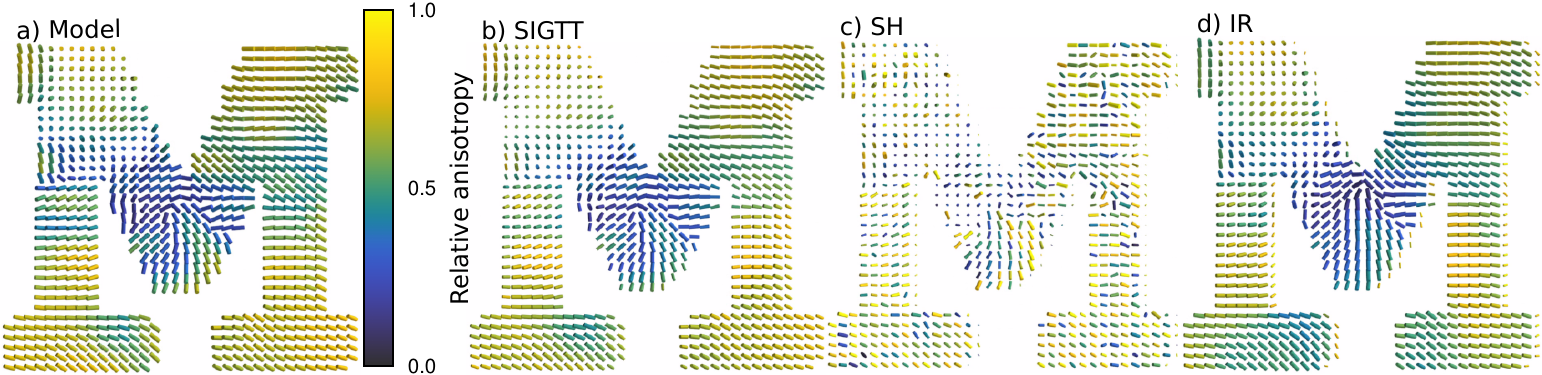}
  \caption{\textbf{Comparison of virtual slices of ``M''.}
    \textbf{a)} Virtual slice of simulated sample ``M''.
    \textbf{b)} \gls{sigtt} reconstruction. 
    \textbf{c)} SH reconstruction.
    \textbf{d)} IR reconstruction.
    The glyphs are colored according to the relative anisotropy (equation~\eqref{eq:aniso}) and scaled according to the isotropic component (equation~\eqref{eq:iso}) of each reciprocal space map. All three reconstructions follow the orientations of the model reasonably well on average. The \gls{sigtt} reconstruction follows both the orientations and relative anisotropy of the model closely. The SH reconstruction has a lot of variation in the relative anisotropy, as well as many orientations deviating from the overall tendency to follow the model. The IR reconstruction follows the model almost  as well as the SIGTT reconstruction.
  }
  \label{fig:m_compare}
\end{figure*}
Volume renders of the errors of each method for ``M'' can be seen in \autoref{fig:m_errors}. 
It is clear from the figure that all methods have larger errors in the middle regions of the model, but that these are larger for SH and IR.
Moreover, outside these regions, the error for \gls{sigtt} is closer to $0$.

In \autoref{fig:m_compare} virtual slices with glyphs showing the orientation of each reciprocal space map for a) the simulation ``M'',\ b) \gls{sigtt},\ c) SH, and d) IR.
See Supplementary note 3 for details on the orientation analysis.
The comparison shows that the orientation of the reciprocal space maps in the simulation are overall reasonably well followed by all reconstructions.
However, from c) it is clear that the SH reconstruction contains many deviations from the simulated orientation.
In terms of the relative anisotropy (equation~\eqref{eq:aniso}), indicated by the color of each glyph, it is generally well followed by \gls{sigtt}, as seen in b), but poorly followed by SH, seen in c).
It can be inferred from this that the numerous deviations in the SH reconstruction are the cause of the large dispersion in $R^2$ seen in \autoref{fig:zonal_and_r2} a).
IR, in d), gives a reconstruction that follows the orientations and relative anisotropy nearly as well as b), with the exception of certain regions.
This is to be expected, since while the anisotropy in fiber-like symmetry is generally dominated by the rank-2 tensor component, this will not be the case everywhere.

\begin{figure}
    \includegraphics[width=\columnwidth]{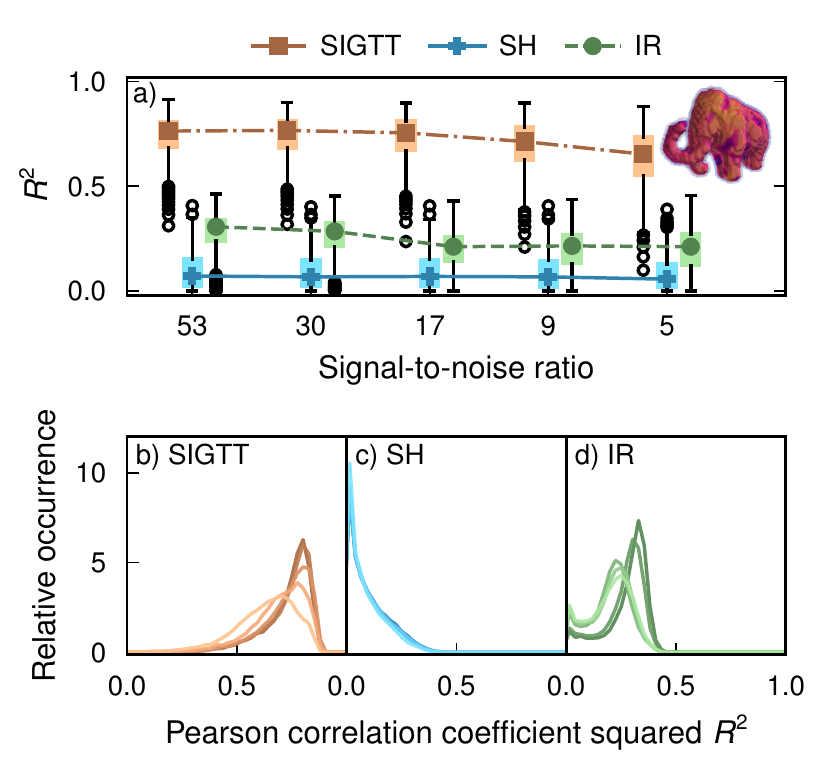}
  \caption{
    \textbf{Correlations for sample ``mammoth''.}
    \textbf{a)} Box plots of $R^2$ as defined in equation~\ref{eq:corr}, with lines and symbols indicating the respective median of each box plot. Outlier dots each represent $100$ voxels. 
    \textbf{b)--d)} Correlation coefficient distribution for each of the three methods. The signal-to-noise ratio goes from 53 (darkest lines) down to to 5 (lightest lines). The image inset shows a volume render of the simulated sample. Panel a) 
  }
  \label{fig:mammoth_stats}
\end{figure}

In panels e)--h) of \autoref{fig:zonal_and_r2} the performance between the three models is compared in reconstructing sample ``T'', which consists of reciprocal space maps with $\ell_\text{max} = 2$.
\gls{sigtt} and IR perform fairly similarly, with the performance of IR consistently being somewhat worse.
Since these models are very similar for the special case of $\ell_\text{max} = 2$, this difference in performance is likely related to the fact that the IR implementation uses a less precise projection algorithm, as well as its use of fixed-stepsize steepest descent gradient optimization.
It is likely the case that \gls{sigtt} performs better due to modeling the measured reciprocal space map as a line integral on the sphere, rather than as a point, as well as using continuity-enforcing regularization.
SH performs very poorly regardless of noise level in the case of ``T''.
This is likely both because it cannot directly represent coefficients of the $\ell = 2$ harmonics due to its use of squared polynomials, and because ``T'' does not follow zonal symmetry.

The results from the reconstruction of the sample ``mammoth'' are seen in \autoref{fig:mammoth_stats}.
\gls{sigtt} has similar performance to the case of sample ``M'', with its correlation starting around $0.8$ and decaying to about $0.65$ as the signal-to-noise ratio decreases, with an increase in the interquartile range.
SH performs very poorly, which is to be expected, as the ``mammoth'' sample does not follow zonal symmetry at all.
IR performs better than SH, but is bounded by the fact that the reciprocal space maps in this simulation only have a weak rank-2 tensor component.
Thus, it does not reach a median $R^2$ above $0.3$.
See supplementary note 4 and supplementary figures 1 and 2 for volume renders of the errors of ``T'' and ``mammoth'', as well as supplementary figure 3 and supplementary equation~(S1) for a comparison of the anisotropic power distribution of ``mammoth`` and ``M''.

\subsection*{Experimental data}
\begin{figure*}
    \includegraphics[width=\linewidth]{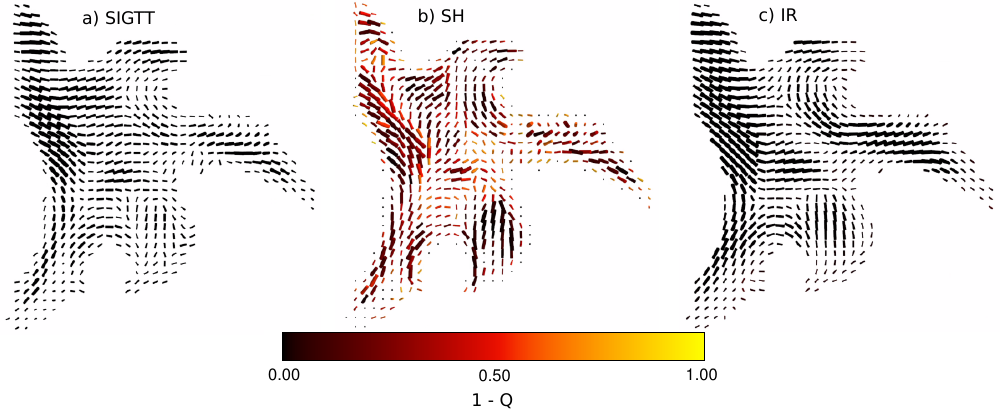}
  \caption{\textbf{Experimental ensemble reconstructions.}
    \textbf{a)} Virtual slice from ensemble of 10 reconstructions each with randomized initial conditions of a sample of trabecular bone using  \gls{sigtt}.
    \textbf{b)} Ensemble reconstruction using SH.
    \textbf{c)} Ensemble reconstruction using IR.
    The glyphs are scaled by the square root of the anisotropic power, defined in equation~\eqref{eq:var}.
    The quantity $Q$, defined in equation~\eqref{eq:aniso_quot}, is a measure of how much the anisotropy of each reciprocal space map changes across the ensemble of reconstructions.
    The methods generally agree in terms of the orientation of each reciprocal space map, but only SH shows a change in the anisotropy across the ensemble.
  }
  \label{fig:trab_compare}
\end{figure*}
An ensemble of 10 reconstructions each with some initial conditions randomized were performed using \gls{sigtt}, SH and IR on a sample of trabecular bone.
For experimental details, see Liebi \textit{et al}. 2015, sample B \cite{liebi_nat_2015}.
The chosen q-range for this reconstruction does not contain the collagen peak, and therefore its reciprocal space map has fiber symmetry.
A comparison of a virtual section from each of the three methods can be seen in \autoref{fig:trab_compare}.
Because there is no ground truth to compare to for experimental data, the ensemble of reconstructions was instead analyzed to investigate the robustness of each method against perturbations in the initial conditions.
The spherical harmonic representations of the 10 reconstructions were averaged over, voxel-by-voxel, and \autoref{fig:trab_compare} shows the results of the averaged reconstruction.
The colors of the orientation glyphs indicate the degree to which the anisotropy of the reciprocal space map changes across every reconstruction; the quantity $Q$ is defined in equation~\eqref{eq:aniso_quot}. 
The glyphs are scaled according to the square root of the mean anisotropic power of each RSM (the anisotropic power is defined in equation~\eqref{eq:var}) across the ensemble.
The results indicate that \gls{sigtt} and IR are robust to perturbations of the initial conditions, but that SH is not.
However, the orientations of the averaged SH reconstruction agree well with those of \gls{sigtt} and IR.
A plausible reason for this difference is that SH is the only method out of the three which depends on Euler angles.
Depending on the initial conditions of the angles, the solution may be confined to local minima, as the symmetries of its reciprocal space can only vary across a limited subspace of the total spherical harmonic coefficient space.
While IR performed similarly to \gls{sigtt} in the chosen q-range, a rank-2 tensor can not contain more than one local maximum per hemisphere.
This poses a problem for the method in the case of the RSM of the collagen peak q-range of bone, which contains two distinct maxima -  one lies along a great circle, and one lies on the poles orthogonal to that great circle.
This symmetry, which requires at least a rank-4 tensor, can be represented by both \gls{sigtt} and SH \cite{manuel_jsr_2020}.
\section*{Discussion}
This work has demonstrated the \gls{sigtt} method for \gls{saxs} tensor tomographic reconstruction of the reciprocal space map in samples using a band-limited spherical function expressed in spherical harmonics (see \nameref{sect:methods} for details).
In three case studies using simulated data with approximately zonally symmetric reciprocal space maps, rank-2 tensors, and complicated-textured higher-order reciprocal space maps, the method produces results superior to the approaches of Liebi \textit{et al.}\ (2018) and Gao \textit{et al.}\ (2019)\cite{liebi_aca_2018, gao_aca_2019}.
\gls{sigtt} has also been shown to be robust to perturbations in the initial conditions when reconstructing experimental data.
The reconstruction of the reciprocal space map up to higher spherical harmonic orders will enable the use of more specific methods of characterization that reveal information about the nature of the sample beyond the main orientation or adherence to a specific, predetermined symmetry.
Samples frequently contain multiple domains of various orientations, and the \gls{sigtt} method makes it possible to study not just individual domains, but also the boundaries and transitions between them, including voxels with more than one orientation.
\gls{sigtt} can be applied to more complicated reciprocal space maps, such as those that occur in \gls{saxs} measurements of samples with hexagonal symmetries, or wide-angle X-ray scattering measurements.
This includes reciprocal space maps which are not necessarily well approximated by a rank-2 tensor.
Similarly, \gls{sigtt} should make it easier to reconstruct samples with smaller or less well-organized domains.
In this way, the reconstruction of the reciprocal space map using band-limited spherical functions makes full use of the data obtained from the collection of scanning \gls{saxs} data at multiple angles, and opens up many new venues of analysis.
Finally, the anchoring of \gls{sigtt} in a framework of integral geometry and linear algebra highlights the potential for algorithms employing alternative schemes both for data acquisition and representation of the reciprocal space map.

\section*{Methods}\setcurrentname{Methods}\label{sect:methods}
\subsection*{Discretized formalism}
\label{sect:formalism}
The inverse problem to the forward model of equation~\eqref{eq:single_measurement} is to obtain the distributions $a_m^\ell(\mathbf{r})$ for a set of measured data.

In practical terms, the solution to the inverse problem is best discussed using a discrete formalism.
Prior to reconstruction, measurements at a particular value of $\mathbf{q}$ are reduced by binning the measured pixel intensities into azimuthal segments, which corresponds to an integral over a line segment on the sphere.
Consequently, for the bin $i$, centered on $\varphi_i$ of width $\Delta\varphi$, the spherical harmonics in equation~\eqref{eq:single_measurement} are integrated to give the coefficients for these bins,

\begin{align}
\overline{Y}_m^\ell(\overline{\upsilon}, i) &= \frac{1}{\Delta \varphi}\int\displaylimits^{\varphi_i + 0.5\Delta\varphi}_{\varphi_i - 0.5\Delta\varphi}\hat{Y}^\ell_m(C(\tau,\upsilon_3,\upsilon_4)) d\tau.
\label{eq:sph_int}
\end{align}

with $\upsilon_i$ as in equation~\eqref{eq:upsilon}.
Since this effectively blurs the reciprocal space map, it constrains the maximum frequency that can be uniquely represented in the reduced data. By representing the great circle cut of a spherical harmonic as a trigonometric polynomial, it can be concluded that the $\ell_\text{max}$ of the fitted harmonic should follow

\begin{align*}
    \ell_{\text{max}} \leq N-1,
\end{align*}

where $N$ is the number of azimuthal bins for $\varphi_i \in [0, \pi)$, due to the assumption of Friedel symmetry.

The sample volume, spanned by $\mathbf{r}$ in equation~\eqref{eq:voxel_model}, is divided into cubic voxels, of the same size as the step size in the scanning \gls{saxs} measurement.
To pose our problem in matrix form, we describe our discretized system as a matrix $\mathbf{X}$ of $N$ rows and $M$ columns, where each row corresponds to a voxel, and each column corresponds to a spherical harmonic coefficient.
Similarly, the measured data is described by an $I \times J$ matrix which we label $\mathbf{D}$, with $I$ being the number of scanned points and $J$ the number of detector segments across all rotation and tilt configurations.
For simplicity of representation, we consider scans and detector segments at different rotations and tilts to be distinct, so that $\mathbf{D}$ takes a sparse block matrix form.
The discrete equivalent of the projection operation $\mathbb{P}\cdot$ (see equation~\eqref{eq:johntransform}), considered across all measured projections, is then given by a sparse $I \times N$ matrix $\mathbf{P}$, describing a mapping between weighted sums of the $N$ voxels to the $I$ scanned points of the sample.
Finally, the mapping from the spherical harmonic representation of the reciprocal space map to the segment-wise detector data is given by an $M \times J$ matrix $\mathbf{Y}$, consisting of the $M$ coefficients calculated in equation~\eqref{eq:sph_int} for each of the $J$ detector segments.

This gives us the system of equations

\begin{align*}
    \mathbf{PXY} = \mathbf{D},
\end{align*}

and we write the solution as the optimization of the least-squares problem

\begin{align}
\mathbf{X}_\text{opt} = \argmin_\mathbf{X}\Vert \mathbf{P}\mathbf{X}\mathbf{Y} - \mathbf{D}\Vert^2.
\label{eq:lsq_form}
\end{align}

See supplementary note 5 for details on how to solve this system with an iterative algorithm.

\subsection*{Reciprocal space map evaluations}
In tomography, the solution is generally assumed to be a reasonably smooth function, and we impose a regularization term on equation~\eqref{eq:lsq_form} to ensure this \cite{nw_math_2001}.
Since we do not want the evaluation or comparison of RSMs (defined by Eqs.~(\ref{eq:reciprocal_space_map},~\ref{eq:voxel_model})) to depend on our choice of coordinate system, it is necessary to use rotational invariants.
A general rotational invariant is the canonical inner product of the spherical harmonics, known as the cross-spectrum function,

\begin{align}
    S_\ell(g,h) = \sum_{m=-\ell}^{\ell}\mathcal{N}(\ell)\mathop{g^{\ell}_m h_m^\ell},
    \label{eq:cross_spectrum}
\end{align}

where $\mathcal{N}(\ell)$ is a normalization factor that depends on the choice of spherical harmonic representation, $\ell$ is the cross-spectrum order, and $g$ and $h$ are two spherical functions \cite{shtools_2018}.
$g^\ell_m$ and $h^\ell_m$ are the coefficients of the spherical harmonic representation of $g$ and $h$, given by

\begin{align*}
g_m^\ell = \int \mathop{d\Omega}
\left[
    g(\theta, \phi) \hat{Y}_m^\ell(\theta, \phi)
\right]
\end{align*}

This discussion is therefore applicable to the analysis of spherical functions of any type, but in the particular case of \gls{saxstt}, $f$ and $g$ are reciprocal space maps as defined by Eqs.~(\ref{eq:reciprocal_space_map},~\ref{eq:voxel_model}).
To regularize the problem by imposing a smoothness condition, we compute a nearest-neighbor similarity term,

\begin{align*}
    (\Lambda g)_{ij} = \lambda\sum_{\ell = 0}^{\ell_\text{max}} \left[S_\ell(g_i, g_i) + S_\ell(g_j, g_j) - 2S_\ell(g_i, g_j)\right],
\end{align*}

where the set of all $(i, j)$ indicates neighboring pairs of voxels, $g_i$ is the spherical function associated with each voxel, and $\lambda$ is a regularization coefficient.
In spherical harmonic coefficient space, this reduces to the squared discrete Laplacian operator on our system matrix weighted by $\lambda$,

\begin{align}
    \Lambda X = \lambda \big(\nabla^2 \mathbf{X}\big)^2.
    \label{eq:laplacian}
\end{align}

Minimizing this term results in maximizing the covariance between neighboring voxels, since

\begin{align}
    \sum_{\ell=1}^{\infty}S_\ell(g,h) = \text{cov}(g,h),
    \label{eq:cov}
\end{align}

with $S_\ell(g, h)$ defined in equation~\eqref{eq:cross_spectrum}, and therefore we also have

\begin{align}
    \sum_{\ell=1}^{\infty}S_\ell(g,g) = \text{var}(g),
    \label{eq:var}
\end{align}

where in particular, $\text{var}(g)$ is called the \textit{anisotropic power} of the RSM represented by $g$.

Thus, with the addition of the regularization term in equation~\eqref{eq:laplacian}, the solution becomes

\begin{align}
\mathbf{X}_\text{opt} = \argmin_\mathbf{X}\big[\Vert \mathbf{PXY} - \mathbf{D}\Vert^2 + \lambda \Vert \nabla^2 \mathbf{X}\Vert^2 \big].
\label{eq:lsq_form_reg}
\end{align}

The \emph{isotropic component} of an RSM is defined as its spherical mean $\overline{\mu}$, and in spherical harmonic representation,

\begin{align}
    \label{eq:iso}
    \overline{\mu}(g) = g^0_0,
\end{align}

where $g$ is the spherical function representing the RSM.

We also define the \emph{relative anisotropy} of a reciprocal space map as

\begin{align}
    \label{eq:aniso}
    \varsigma(g) = \frac{\sqrt{\text{var}(g)}}{g^0_0} = \frac{\sigma(g)}{\overline{\mu}(g)},
\end{align}

or in other words, the standard deviation $\sigma(g)$ of the spherical function $g$ normalized by the spherical mean $\overline{\mu}(g)$.
This is useful to indicate the texture of a sample in many-voxel visualizations, and is used to color the glyphs in \autoref{fig:m_compare}.
While this quantity does not have an upper bound and does not take symmetry into account, it has the advantage of being easily calculated for any reciprocal space map model irrespective of basis or normalization.
This is similar but not identical to the quantity referred to as \textit{degree of orientation} in Liebi \textit{et al.}\, (2018, equation (8)) \cite{liebi_aca_2018}, which evaluates to the variance of the square root of the reconstructed reciprocal space map divided by its mean.


\subsection*{Correlation calculations}
Because the variances and covariances of reciprocal space maps (Eqs.~(\ref{eq:cov},~\ref{eq:var})) are rotational invariants, compositions of them are also invariants, and in particular

\begin{align}
    \label{eq:corr}
    \frac{\text{cov}(g,h) ^ 2}{\text{var}(g)\text{var}(h)} = R^2(g, h),
\end{align}

where $\text{var}(g)$ is the variance of the spherical function $g$, $\text{cov}(g,h)$ is the covariance of two spherical functions, and $R^2(g,h)$ is the squared Pearson correlation coefficient of the two functions $g, h$.
The squared Pearson correlation coefficient is used in the comparison of reconstructed reciprocal space maps to those in simulated samples.
In the calculation of $R^2$ shown in Figs.~\ref{fig:zonal_and_r2} and \ref{fig:mammoth_stats}, the calculation is done between the reciprocal space maps of each voxel in the simulated model (excluding empty voxels) and the reciprocal space map of the same voxel in the reconstruction.
The box plots in \autoref{fig:zonal_and_r2} a) and e) and \autoref{fig:mammoth_stats} a) follow the original definitions of Tukey (1970) \cite{tukey77}.
They are defined such that the colored rectangles span the interquartile range of the correlation distribution.
The black ``whiskers'' outside the colored rectangles span the smallest and largest value in the range $[Q_1 - 1.25\cdot (Q_3 - Q_1), Q_3 + 1.25\cdot(Q_3 - Q_1)]$, where $Q_i$ is the $i$th quartile of the distribution of $R^2$.
Values outside the range of the ``whiskers'' are represented by small circles, with each circle showing the mean $R^2$ of $100$ reciprocal space maps.
If there is at least 1, but fewer than 100 reciprocal space maps above or below each whisker, a single black circle is shown, representing the mean of $R^2$ across these reciprocal space maps.
The median, equivalent to $Q_2$, is shown by the colored markers in each box plot.

\subsection*{Simulation framework}

For each of the three sample volumes, source points were determined such that the distance between each point was maximized; 4 source points for ``M'', 2 source points for ``T'', and 5 source points for ``mammoth''.
Band-limited spherical functions were constructed such that the spectral power of each order followed power-law decays with respect to $\ell$, and assigned to each source point.
The interior distance from each source to every other point in the volume was then approximately computed using a combination of a k-d tree, and Dijkstra's algorithm.
The sources were assigned correlation lengths, with the assumption that for each order of each spherical function in the sample, correlation with the source would decay with distance like a Gaussian distribution with the correlation length as its standard deviation.
The remaining spherical functions were then solved for under several constraints - in all cases, it was assumed that the spherical functions of neighboring voxels in the volume would be correlated with each other, and that the power of each order of the function in each voxel would equal a distance-weighted average of the power of its source.
Moreover, it was required that all functions be non-negative.
Non-negativity is difficult to enforce perfectly for spherical polynomials, but a dense sampling of each function was performed and the isotropic component was increased to eliminate all the detected negative points.


``M'' consists of spherical polynomials up to $\ell_\text{max} = 12$ that approximately follow zonal symmetry.
To enforce zonal symmetry, each spherical function was required to correlate with the $\ell$-weighted spherical harmonic Dirac $\delta$ function

\begin{align*}
    w(\ell) \delta_\ell^m(\alpha, \beta) = (-1)^\frac{\ell}{2} Y_\ell^m(\alpha, \beta).
\end{align*}

with $(\alpha, \beta)$ given by the fiber-like orientation of the $\ell = 2$ component of the reciprocal space map, and $w(\ell)$ being an $\ell$-weighting function.
In general, the condition of zonal symmetry requires compromise with the demand of continuity (enforced by minimizing the spherical harmonic Laplacian, as in equation~\eqref{eq:laplacian}).
This is because the different orders of spherical harmonics have differing symmetries with respect to rotations.
In effect, this leads to some attenuation of parts of the great circle of intensity.
Hence, the reciprocal space maps of ```M'' are said to only approximately follow zonal symmetry.
``T'' is entirely represented by symmetric rank-2 tensors, with a distribution required to be continuous.
``mammoth'' is represented by spherical polynomials with $\ell_\text{max} = 8$, which in addition to being continuous have a dampened $\ell = 2$ component, such that the $\ell = 2$ and $\ell = 4$ components have approximately the same spectral power.
This was done to approximate the type of spectral power distributions that occur in reciprocal space maps which do not either have fiber-like or peak-like symmetries, such as the collagen peak in bone.
The reciprocal space map cuts generated from the projections of each simulation where divided into eight segments, accounting for Friedel symmetry, and integrated over, emulating the azimuthal integration approach used for experimental data.
The choice of eight bins was chosen based on previous usage in experimental data, as well as the fact that it will restrict the bandlimit of spherical functions that can be precisely retrieved to $\ell_\text{max} = 6$, meaning that it will not be possible for \gls{sigtt} to exactly solve for the reciprocal space maps of the samples ``M'' ($\ell_\text{max} = 12$) and ``mammoth'' ($\ell_\text{max} = 8$).

\subsection*{Ensemble reconstructions}
For the ensemble reconstructions shown in \autoref{fig:trab_compare}, each of the methods had their initial conditions randomized.
In the case of \gls{sigtt} and IR, this consisted of randomizing the coefficients of their solution vectors at values several orders of magnitude below what is expected from their reconstructed value.
In the case of SH, the randomization was only applied to the Euler angles of the orientations of each voxels, which must be initialized before each reconstruction.
The angles were randomized such that the orientations would be uniformly distributed on the sphere.
The stepwise reconstruction procedure of \cite{liebi_aca_2018} was then followed.
In order to average over the result of the ensemble, the squared coefficients of the SH reconstruction, performed with $\ell_\text{max} = 6$, were expanded through Driscoll-Healy quadrature in a non-squared spherical harmonic representation up to $\ell_\text{max} = 12$ \cite{driscoll_healy_1994}.
Tests incorporating higher orders and denser grids showed that this approach was accurate to a relative error of approximately $0.1\%$ in the variance of the reciprocal space map, which was deemed sufficient for the purpose of examining the reconstruction's consistency across the ensemble.
To evaluate the robustness of the reconstructions with respect to initial conditions, we calculated an anisotropic power quotient for the reciprocal space map in each voxel,

\begin{align}
    \label{eq:aniso_quot}
    Q = \frac{\text{var}(\frac{1}{n}\sum_{i=0}^{n}g_i)}{\frac{1}{n}\sum_{i=0}^n \text{var}(g_i)},
\end{align}

where $g_i$ is the spherical function representing the reciprocal space map in each voxel for ensemble reconstruction $i$, and $\text{var}(g_i)$ is the anisotropic power of the reciprocal space map as defined in equation~\eqref{eq:var}, for $n$ total reconstructions.
If the reciprocal space maps of every reconstruction in the ensemble are identical, the value of $Q$ will be $1$, and it will be in the range $[0, 1)$ if the reciprocal space maps differ.

\subsection*{Implementations}
\label{sect:implementations}
\gls{sigtt}, and the simulations used in this work was implemented in Python.
The most demanding parts of the code, projections and back-projections, are carried out using Numba, and is part of the software package Mumott, whereas other calculations are carried out in NumPy and SciPy \cite{numpy, scipy, lam_2015_numba}.
Mumott is available at \href{http://doi.org/10.5281/zenodo.7798530}{http://doi.org/10.5281/zenodo.7798530}.
The projection code, written specifically for this work, uses only CPU resources.
It performs the John transform by using vectors to trace out the lines of integration, and sampling the voxels that these intersect with, in proportion to the lengths of the intersecting segments.
It is likely that a considerable speedup could be obtained by using GPU-based projection.
Visualizations of the three-dimensional reconstructions were created using the Python package \textsc{Mayavi} \cite{mayavi}.
The color maps used throughout this paper were generated with the help of ColorCET (\url{http://colorcet.com}) \cite{kovesi_2015}.
All computations were performed on a workstation using a 12-core, 4.6 GHz AMD Ryzen 9 3900X CPU, and 64 GB DDR4 2666 MHz RAM.
For the IR and SH methods, the original code from the cSAXS software package written in Matlab was used, with modifications for optimization and termination of each reconstruction upon convergence.
The projection code in this package samples voxels using coordinate transforms and bilinear interpolation.
It slices the sample along the plane of integration, and samples the four voxels closest to the line of integration, based on the distance in the plane of projection.
Because it effectively treats the projection of each voxel as a square at every angle of projection, and does not consider the full three-dimensional distance between voxels, this approach suffers from high-frequency artefacts.
However, following testing, this approach was deemed sufficiently accurate for the purpose of comparing SH and IR to \gls{sigtt}.
See supplementary note 6 and supplementary figure 4 for timing comparisons between the methods.

\section*{Data availibility}
The simulated data created for and used in this work is available at \href{http://doi.org/10.5281/zenodo.7673985}{https://doi.org/10.5281/zenodo.7673985}.
A notebook demonstrating the analysis and reconstruction using Mumott is available at \href{http://doi.org/10.5281/zenodo.7799517}{http://doi.org/10.5281/zenodo.7799517}.

\section*{Acknowledgments}
This work was funded by the Swedish research council (VR 2018- 041449) and the European research council (ERC-2020-StG 949301).
Some of the computations in this work were enabled by resources provided by the Swedish National Infrastructure for Computing (SNIC) at NSC, C3SE and PDC partially funded by the Swedish Research Council through grant agreement no. 2018-05973.
The authors would also like to extend their gratitude to William Lionheart at the University of Manchester for providing input and feedback on spherical harmonic and tensor field formalism.
We acknowledge the Paul Scherrer Institute, Villigen, Switzerland for provision of synchrotron radiation beamtime at the beamline cSAXS of the SLS.

\section*{Author contributions}

L.C.N.: Writing, editing, methodology, author of novel code, figure creation, method development, analysis. P.E.: Coding support, algorithm suggestion, methodology, supervision, reviewing, and editing. M.G-S.: Methodology, data curation, experimental background, co-author of previously published code, reviewing, and editing. M.L.: Conceptualization, methodology, supervision, experimental background, co-author of previously published code, analysis, reviewing, and editing.



\section*{Competing interests}

The authors declare no competing financial or non-financial interest.

\bibliography{lit.bib}

\end{document}


\title{
    Supplementary Information: \texorpdfstring{\\}{}
    Small-angle scattering tensor tomography algorithm \texorpdfstring{\\}{}
    for robust reconstruction of complex textures
}

\author{Leonard C. Nielsen}
\email{leoniel@chalmers.se}
\affiliation{\chalmersphys}
\author{Paul Erhart}
\affiliation{\chalmersphys}
\author{Manuel Guizar-Sicairos}
\affiliation{\affilpsi}
\affiliation{\affilepfl}
\author{Marianne Liebi}
\email[Corresponding author.]{marianne.liebi@psi.ch}
\affiliation{\chalmersphys}
\affiliation{\affilpsi}
\affiliation{\affilepfl}
\maketitle
\begin{description}
  \item[Supplementary Note 1] Coordinate mapping
  \item[Supplementary Note 2] Tensors and spherical harmonics
  \item[Supplementary Note 2] Orientation analysis
  \item[Supplementary Note 4] Additional results
  \item[Supplementary Note 5] Solving regularized least-squares system
  \item[Supplementary Note 6] Implementation details and timing
\end{description}
\section{Coordinate mapping}
In the ``Formalism'' section of the paper, the John Transform is described using vectors in the local coordinate system of the sample.
This is the most convenient way to parameterize the system, since it is the sample volume that is ultimately being solved for.
However, for practical reasons, it is useful to lay out how one can map a coordinate system of a laboratory frame, relative to which the sample is moving.
In general, we can specify the relationship between the laboratory system and the sample system by choosing a set of vectors and rotation axes at zero tilt and rotation, and then applying any rotations of the sample to those vectors.
Three positioning vectors are needed - one for the direction of the x-ray beam, and two which indicate how the beam moves relative to the sample during scanning.
We may call these $\mathbf{p}$, $\mathbf{w}_j$ and $\mathbf{w}_k$.
The vectors $\mathbf{w}_j$ and $\mathbf{w}_k$ should be shifted to align the system, such that if $\mathbf{w}_j = \mathbf{w}_k = 0$, the beam intersects the center of the sample at all rotations.
Additionally, two are required to indicate how angles on the detector map to the sphere, we may call these $\mathbf{q}_0$ and $\mathbf{q}_{90}$, defined to indicate where on the sphere detector angle $0$ and $90$ lie.
Two orthogonal rotation axes need to be defined, orthogonal to the projection direction, in order to translate rotations to the sphere of projection, which we call $\mathbf{\hat{\alpha}}$ and $\mathbf{\hat{\beta}}$.
These do not necessarily have to correspond to the axes of real rotation stages.
Then, any rotation $\mathbf{R}$ that the sample is subject to must be decomposed into three components, $\mathbf{R}_\alpha(\alpha)$, $\mathbf{R}_\beta(\beta)$, and $\mathbf{R}_p(\gamma)$, for rotations around $\mathbf{\hat{\alpha}}$, $\mathbf{\hat{\beta}}$ and $\mathbf{p}$ respectively, such that

\begin{align*}
    \mathbf{R} = \mathbf{R}_p(\gamma)\mathbf{R}_\beta(\beta)\mathbf{R}_\alpha(\alpha)
\end{align*}

Then, any one of the vectors $\mathbf{a}$ in the zero-rotation system will become

\begin{align*}
    \mathbf{a}' = \mathbf{R}^T_\alpha(\alpha)\mathbf{R}^T_\beta(\beta)\mathbf{R}^T_p(\gamma)\mathbf{a}
\end{align*}

The John transform parameters $j$ and $k$ are then given by the norms of $\mathbf{w}_j$ and $\mathbf{w}_k$, whereas $\alpha$ and $\beta$ are the arguments of the rotation matrices $\mathbf{R}_\alpha$ and $\mathbf{R}_\beta$.
The rotation matrix $\mathbf{R}_p$ simply specifies a rotation around the axis of projection, and therefore only changes the mapping of $j$ and $k$, rather than representing a degree of freedom on the sphere of projection.
The reciprocal space angles $\theta$ and $\phi$ are given by the azimuthal and polar angle of the reciprocal space vector
\begin{align*}
    \mathbf{q}' = \mathbf{R}^T(\cos(\varphi)\mathbf{q}_0 + \sin(\varphi)\mathbf{q}_{90})
\end{align*}
where $\varphi$ is the angle on the detector.

The vectors $\mathbf{v}$ and $\mathbf{u}$ in the John transform are given by

\begin{align*}
    \mathbf{v} &= \mathbf{R}^T(\mathbf{w_j} + \mathbf{w_k}) \\
    \mathbf{u} &= \mathbf{R}^T(\mathbf{p}) \\
\end{align*}

Using these rules, one can calculate the John transform using rotation information from any laboratory coordinate system.

\section{Tensors and spherical harmonics}

The comparison between methods using spherical harmonic representations and tensor representation merits a brief discussion of isomorphisms between these parameterizations.
Spherical harmonics are simply harmonic polynomials on the sphere.
In this work we use the \emph{real} spherical harmonics, which we may define as

\begin{align*}
    \hat{Y}^\ell_m(\theta, \phi) = \begin{cases}
    L^\ell_m(\theta) \cos(m  \phi) &\text{$m \geq 0$}\\
    L^\ell_m(\theta) \sin(\vert m \vert \phi) &\text{$m < 0$}
    \end{cases}
\end{align*}

where $\theta$ is the polar angle, $\phi$ is the azimuthal angle, $L^\ell_m(\theta)$ is the associated Legendre polynomial of degree $\ell$ and order $m$, with $\ell \geq 0$ and $\vert m \vert \leq \ell$.
Because of this, they have an intuitive mapping to traceless symmetric Cartesian tensors when representing functions on the unit sphere.
In particular, a band-limited spherical function represented in $\ell = n$ maps directly to a traceless rank-$n$ tensor.
The rank-0 tensor is simply a constant, so it naturally maps to $\hat{Y}^0_0(\theta, \phi)$, which is a constant term.
This mapping naturally leads to restricting the mapping to \emph{traceless} tensors, because the trace of a tensor on the unit sphere is a constant term.
Moreover, a spherical harmonic of $\ell = n$ can be represented as a polynomial of Cartesian coordinates of degree $n$, that is to say

\begin{align*}
    P(n, \mathbf{x}) &= a_m\hat{Y}^n_m(\mathbf{x}) \\
    x_ix_i &= 1
\end{align*}

where Einstein summation over repeated indices is used, and $\mathbf{a}$ is some vector of $2n + 1$ coefficients. In the same way, we have for a traceless rank-$n$ tensor $T$ the mapping

\begin{align*}
    P(n, \mathbf{x}) = T^{ijk\ldots}x_ix_jx_k\ldots,
\end{align*}

$P(n, \mathbf{x})$ is the value of some polynomial of degree $n$ at $\mathbf{x}$.
Restricting ourselves to the unit sphere, it is then the case that

\begin{align*}
    a_m\hat{Y}^n_m(\mathbf{x}) = T^{ijk\ldots}x_ix_jx_k\ldots,
\end{align*}

By the additivity of polynomials, this mapping can be extended to sums of traceless symmetric tensors tensors of arbitrary orders, and through Gaussian elimination, it is straightforward to calculate a mapping between spherical harmonics and symmetric Cartesian tensors.
Tensor and spherical harmonic representations of functions on the unit sphere are thus isomorphic, and their different analytical and algebraic properties render them suitable for different purposes.
Spherical harmonics are especially suitable for calculations where rotational invariance is crucial, and for calculating spherical statistics such as the variance or mean over the sphere.
On the other hand, it would be easier to perform tensor calculus (such as computing the divergence or curl) on Cartesian tensors.

\section{Orientation analysis}
Orientation analysis is used in figures 4 and 6 of the main work, as well as in the creation of the simulated sample ``M''.
The identification of a reciprocal space map's orientation is done by eigenvector-eigenvalue decomposition of its $\ell = 2$ spherical harmonic coefficients.
The $\ell = 2$ coefficients are translated into a symmetric traceless rank-2 tensor by solving for each coefficient in the tensor's polynomial representation.
Specifically, on the unit sphere, the two representations may be expanded as

\begin{align*}
    \begin{bmatrix}x && y && z \\  \end{bmatrix} \begin{bmatrix}
    T_{xx} && T_{xy} && T_{xz} \\
    T_{xy} && T_{yy} && T_{yz} \\
    T_{xz} && T_{zy} && T_{zz} \\
    \end{bmatrix} \begin{bmatrix}x \\ y \\ z\end{bmatrix} &= x^2T_{xx}  +  y^2T_{yy} + z^2T_{zz}  + 2(xyT_{xy}  + xzT_{xz}  + yzT_{yz} ) \\
    \mathcal{N}(2)\begin{bmatrix} a_{0}^2 && a_{1}^2 && a_{-1}^2 && a_{2}^2 && a_{-2}^2 \end{bmatrix} \begin{bmatrix} \frac{2 z^2 - y^2 - x^2}{2 \sqrt{3}} \\ xz \\ yz \\ xy \\ \frac{x^2 - y^2}{2} \end{bmatrix} &= \mathcal{N}(2)(x^2(\frac{a_{-2}^2}{2}-\frac{a_{0}^2}{2\sqrt{3}}) + y^2(-\frac{a_{-2}^2}{2}-\frac{a_{0}^2}{2\sqrt{3}}) + z^2\frac{a_{0}^2}{\sqrt{3}} + a_{2}^2 xy + a_{1}^2 xz + a_{-1}^2 yz)
\end{align*}
where $\mathcal{N}(2)$ is a normalization factor depending on the spherical harmonic representation, $a_m^2$ is a spherical harmonic coefficient with $\ell$ = 2, $T_{ii}$ are components of a rank-2 tensor, and $(x, y, z)$ are cartesian coordinates.
By equating these two systems of equations, we can then identify each $T_{ii}$ with a linear combination of coefficients $a_m^2$ and thus construct a rank-2 tensor out of spherical harmonic coefficients.
Subsequently, we may solve the eigenvalue problem for the rank-2 tensor, to obtain three eigenvalues (which will sum to zero as the tensor will be traceless) and three orthogonal eigenvectors.
For a rank-2 tensor with three distinct eigenvalues, the eigenvectors associated with the largest and smallest eigenvalues correspond to the location of the minima and maxima of the spherical function.
The eigenvector of the central eigenvalue corresponds to a saddle point.
If there is degeneracy in the eigenvalues, but they are nonzero, the unique eigenvalue corresponds to one extremum (minimum if it is negative, maximum if it is positive), whereas the two identical eigenvalues correspond to two orthogonal points on a set of extreme points which lie along a great circle.
The orientation analysis then depends on assumptions about the symmetry of the nanostructure associated with the reciprocal space map.
If the nanostructure is known or assumed to be scattering from fiber-like structures, orthogonal to the direction of the fibers, the eigenvector associated with the smallest eigenvalue is taken to define the orientation.
On the other hand, if the reciprocal space map is known or assumed to be scattering along the orientation, the eigenvector associated with the largest eigenvalue is taken to define the orientation.
Nanostructures which contain both of these scattering tendencies, such as bone around the q-range of the collagen peak, cannot in general be robustly analyzed in this manner, because each scattering tendency will tend to cancel out the contribution from the other to the rank-2 tensor component of the reciprocal space map.
Such reciprocal space maps require a different treatment, such as correlation analysis with an ensemble of model functions.
\section{Additional results}
This section shows some additional results for the simulation comparisons to supplement those in the main work.
\begin{figure}
    \centering
    \includegraphics[width=\linewidth]{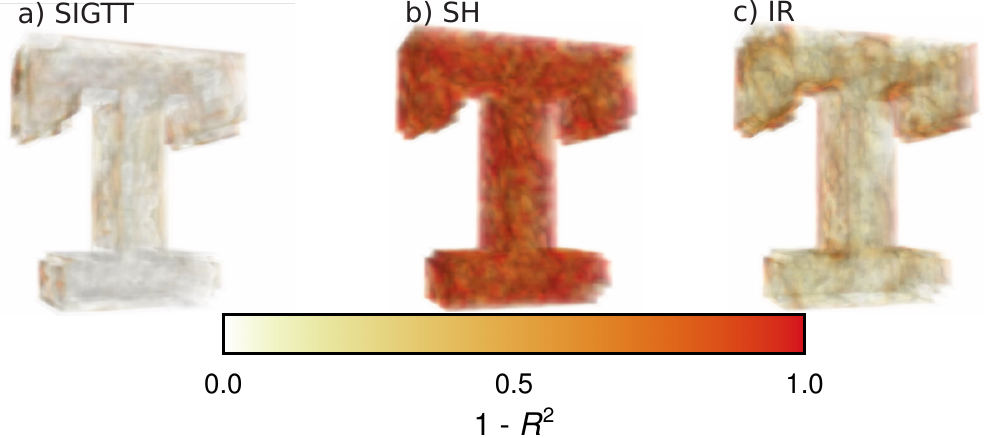}
    \caption{\textbf{Errors for ``T''.} Volume renders of the errors of a) SIGTT, b) SH, and c) IR for sample ``T'', defined as $1 - R^2$, where $R^2$ is given by equation~(15) of the main text. Larger errors are rendered with greater opacity and are thus visible even if they are in the interior.}
    \label{fig:t_errors}
\end{figure}

Volume renders of the errors of each method for ``T'' can be seen in \autoref{fig:t_errors}.
It is evident from the figure that \gls{sigtt} has the smallest errors, and tha the large errors are concentrated around the edges.
It is followed by IR, which has large regions of small errors but larger errors around the same regions as \gls{sigtt}
However, the errors for SH are far greater than for either of the other two methods.
This is partly because the squared polynomials of SH cannot exactly represent rank-2 tensors, and partly because sample ``T'' is not constrained to zonal symmetry.

\begin{figure}
    \centering
    \includegraphics[width=\linewidth]{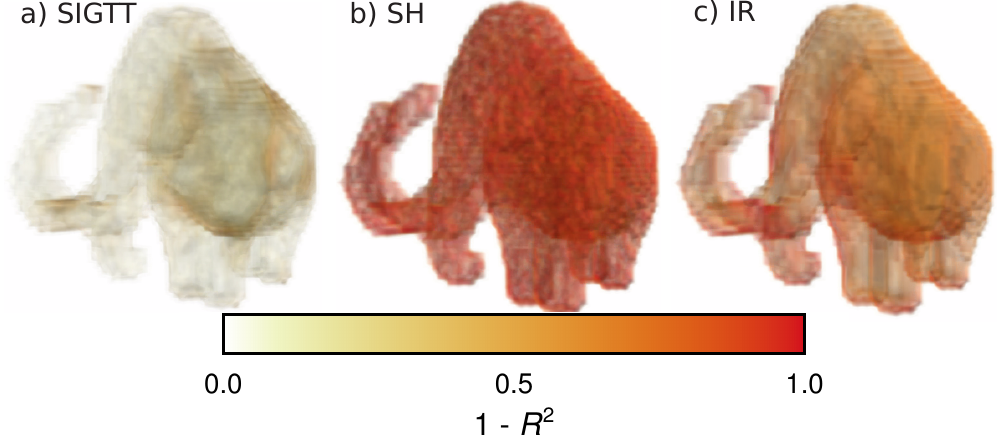}
    \caption{\textbf{Errors for ``mammoth''.} Volume renders of the errors of of a) SIGTT, b) SH, and c) IR for sample ``mammoth'', defined as $1 - R^2$, where $R^2$ is given by equation~(15) of the main text. Larger errors are rendered with greater opacity and are thus visible even if they are in the interior.}
    \label{fig:mammoth_errors}
\end{figure}

A similar set of volume renders of errors for each method for ``mammoth'' can be seen in \autoref{fig:mammoth_errors}.
Again, \gls{sigtt} has the smallest errors, with larger errors around some edge regions.
In this case, SH and IR are more similar, having large errors throughout, but IR consistently appears to have somewhat smaller errors.
This is consistent with figure 5 in the main text. where is it clear that IR can achieve a correlation around $0.3$, but not more, whereas the correlation for SH is close to $0$.
These large errors for IR and SH are due to the fact that the ``mammoth'' sample has a small rank-2 tensor component on the one hand, and on the other hand, no zonal symmetry constraint.

\begin{figure}
    \centering
    \includegraphics[width=0.5\linewidth]{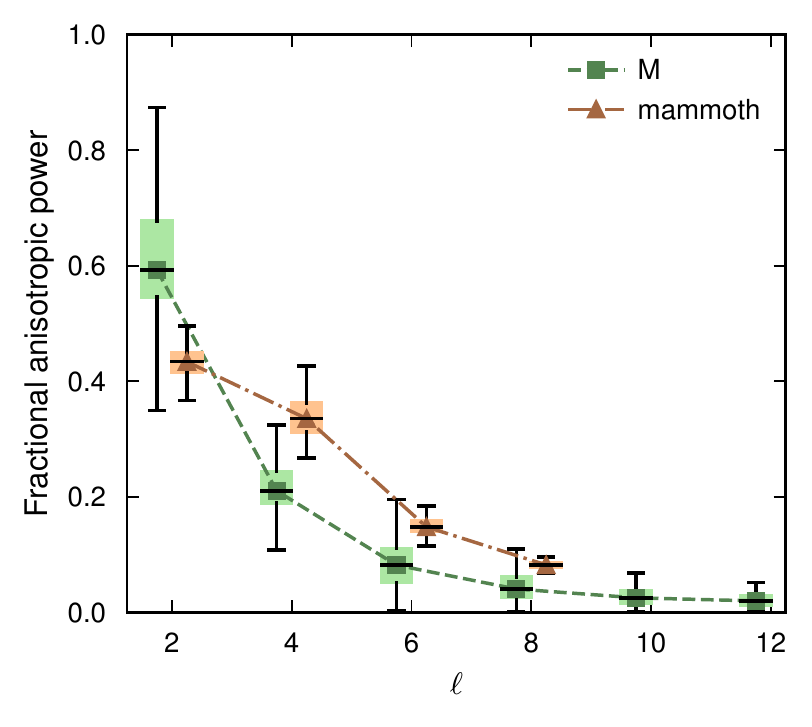}
    \caption{\textbf{Fractional anisotropic power.} Distribution of the fraction of anisotropic power (equation~\eqref{eq:frac_aniso_power}) at each $\ell$ for the samples ``M'' and ``mammoth''.
    Sample ``M'' has a distribution that is more dominated by the $\ell = 2$ component, and a steeper decline, whereas for ``mammoth'', the $\ell = 2$ and $\ell = 4$ components are closer to each other, and $\ell = 6$ and $\ell = 8$ also occupy a larger fraction than for ``M''.}
    \label{fig:power}
\end{figure}

To examine the distribution of anisotropic power in samples, we define the quantity

\begin{align}
    F_\ell(g) = \frac{S_\ell(g)}{\text{var}(g)}
    \label{eq:frac_aniso_power}
\end{align}

where $g$ is a spherical function, and $S_\ell$ and $\text{var}(g)$ are defined in Equations~(8) and (11) in the main text respectively.
This quantity, referred to as the fractional anisotropic power, is then evaluated for every voxel that is part of the sample..
In \autoref{fig:power}, we see box plots of the fractional anisotropic power of ``M'' and ``mammoth''.
While ``M'' has a spectrum that goes up to $\ell = 12$, the orders about $\ell = 6$ only contribute a small amount of the overall power.

\section{Solving regularized least-squares system}

The system of linear equations to be solved in \gls{sigtt} is given as

\begin{align*}
\mathbf{X}_\text{opt} = \argmin_\mathbf{X}\big[\Vert \mathbf{PXY} - \mathbf{D}\Vert^2 + \lambda \Vert \nabla^2 \mathbf{X}\Vert^2 \big].
\end{align*}

This system is solved through the gradient-based method L-BFGS-B.
In detail, a general solution using a quasi-Newton algorithm may be written

\begin{align*}
2r_i &= \Vert \mathbf{PX}_i\mathbf{Y} - \mathbf{D}\Vert^2 + \lambda \Vert \nabla^2 \mathbf{X}_i\Vert^2 \\
\nabla r_i &= \mathbf{P}^T(\mathbf{PX}_i\mathbf{Y} - \mathbf{D})\mathbf{Y}^T + \lambda \nabla^2 \mathbf{X}_i\\
\mathbf{X}_{i+1} &= \mathbf{X}_i- \alpha_i\nabla{r_i} + \beta_i\mathbf{p}_i
\end{align*}

where $r_i$ is the residual for iteration $i$, $\mathbf{X}_i$ is the estimated solution, $\alpha_i$ and $\beta_i$ are method-dependent scalars, and $\mathbf{p}_i$ is a method-dependent momentum term.
While the multiplication with $\mathbf{Y}^T$ is directly implemented as a matrix multiplication, this is not practical to do for the adjoint projection operation $\mathbf{P}^T$, due to the prohibitively large size of the matrix.
Instead, it is computed using the common iterative tomographic method of back-projection of the residual.

\section{Implementation details and timing}

For each of the simulated data sets, an overall signal-to-noise ratio (SNR) was estimated.
Noise was added to each projected simulated data point per

\begin{align*}
    I_{noise} = \frac{\text{poisson}(I \cdot k)}{k}
\end{align*}

with $k$ being a noise parameter, and $\text{poisson(x)}$ being a sampling of $x$ with added Poisson noise.
The values used for $k$ were $k = 10^t$ with $t \in \{2, 1.5, 1, 0.5, 0\}$,
This means that each data point will be a random variable with variance $I / k$.
Thus, the signal-to-noise ratio for a simulated data set with noise parameter $k$ was estimated as

\begin{align*}
    \text{SNR} \approx \sqrt{\overline{\mu}(I_i) k}
\end{align*}

with $\overline{\mu}(I_i)$ being the average intensity of all non-background data points in that data set.
This SNR should be understood principally as a relative measure for a particular simulation, not as an absolute measure.

\begin{figure}
    \centering
    \includegraphics[width=\linewidth]{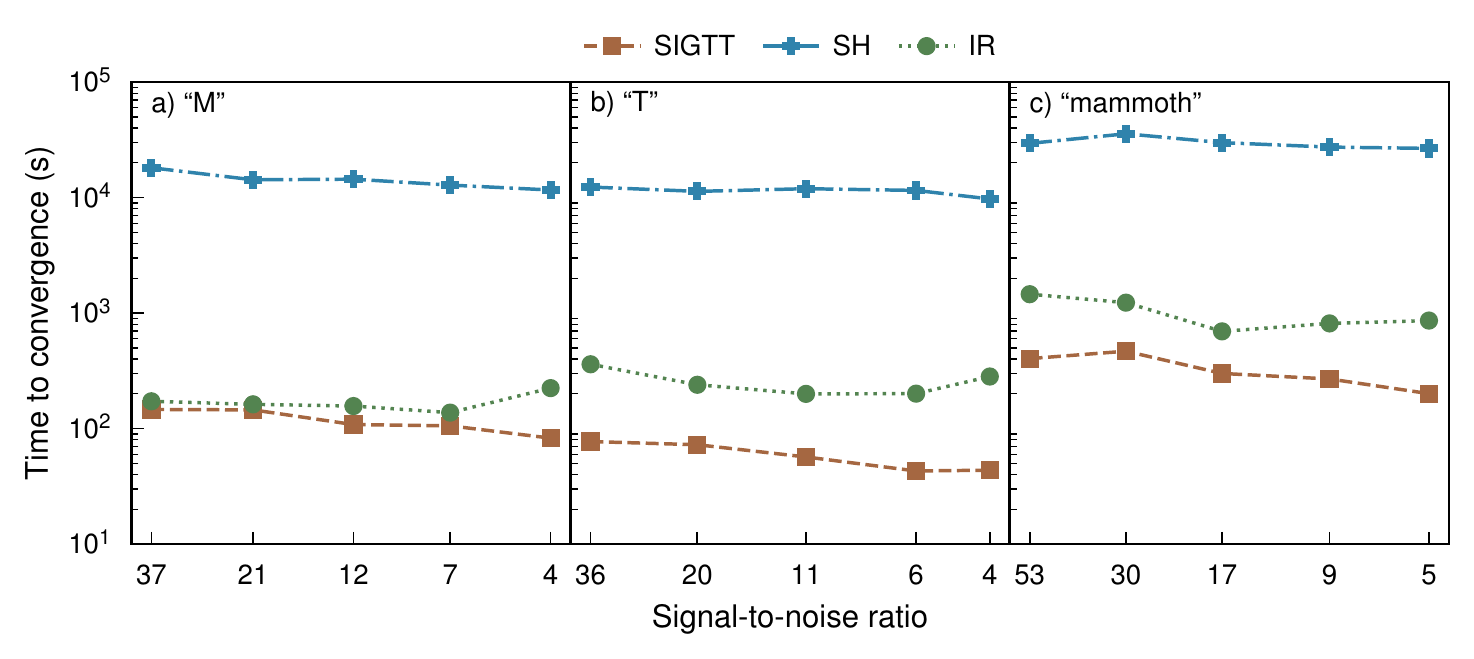}
    \caption{\textbf{Timing comparison of methods.} \textbf{a)} Timing for sample ``M''.
    \textbf{b)} Timing for sample ``T''.
    \textbf{c)} Timing for sample ``mammoth''.
    Note the logarithmic y-axis.
    In all cases, \gls{sigtt} is the fastest method, followed by IR, with SH being the slowest by a considerable margin.}
    \label{fig:timing}
\end{figure}

In \autoref{fig:timing} we show timing information for the reconstructions of  a) ``M'',  b) ``T'', and c) ``mammoth''.
Note the logarithmic y-scale.
In all cases, \gls{sigtt} is the fastest, followed by IR, with SH being significantly slower than either method.
In the case of ``M'' (panel b)), IR is nearly as fast as \gls{sigtt}, although it should be noted that in this case \gls{sigtt} is fitting $28$ coefficients, whereas IR is fitting $6$ coefficients.
For ``T'' (panel b)), both \gls{sigtt} and IR are fitting $6$ coefficients, as ``T'' has only rank-2 tensor components, and in this more directly comparable case, the difference in speed is more substantial.
The sample ``mammoth'' takes substantially longer to fit for all of the methods, as it has a greater volume, viz., 60x60x80 rather than 50x50x50 voxels, which increases the amount of time needed to compute each projection.
This comparison should only be understood as a broad guideline of the performance of the methods; each implementation has a number of parameters which affect convergence in different ways, which were adjusted with the aim of obtaining a reconstruction that correlates well with the simulation, rather than optimized for speed.
Moreover, \gls{sigtt} is implemented in Python, whereas SH and IR are implemented in Matlab, which means that the conditions for optimizations such as multithreading and efficient memory handling are different.
Finally, it is very likely that all methods could be sped up substantially by utilizing a GPU-based projection algorithm.